\documentclass[twocolumn,floatfix,superscriptaddress,amsmath,showpacs,showkeys,aps,prb]{revtex4-1}

\usepackage[utf8]{inputenc}
\usepackage[final]{graphicx}
\usepackage{epsfig}
\usepackage{bm}
\usepackage[normalem]{ulem}
\usepackage[usenames]{color}
\usepackage{times}
\usepackage{verbatim}
\usepackage{xcolor}

\newcommand{\beq}{\begin{equation}}
\newcommand{\eeq}{\end{equation}}
\newcommand{\beqa}{\begin{eqnarray}}
\newcommand{\eeqa}{\end{eqnarray}}

\newcommand{\veck}{\vec{k}}

\newcommand{\ddk}{\frac{d^2 k}{(2\pi)^2}}

\begin{document}
\title{The nematic phase in stripe forming systems within the self consistent screening approximation}
\author{Daniel G. Barci}
\affiliation{Departamento de F{\'\i}sica Te\'orica,
Universidade do Estado do Rio de Janeiro, Rua S\~ao Francisco Xavier 524, 20550-013  
Rio de Janeiro, Brazil}
\author{Alejandro Mendoza-Coto}
\affiliation{Departamento de F\'{\i}sica,
Universidade Federal do Rio Grande do Sul,
CP 15051, 91501-970 Porto Alegre, RS, Brazil}
\author{Daniel A. Stariolo}
\affiliation{Departamento de F\'{\i}sica,
Universidade Federal do Rio Grande do Sul,
CP 15051, 91501-970 Porto Alegre, RS, Brazil}
\date{\today}

\begin{abstract}
We show that in order to describe the isotropic-nematic transition in stripe forming systems with isotropic competing interactions of the Brazovskii class
it is necessary to consider the next to leading order in a $1/N$  approximation  for the effective Hamiltonian. This can be conveniently accomplished 
within the self-consistent screening approximation. We solve the relevant equations and show that the self-energy in this approximation is able to generate 
the essential wave vector dependence to account for the anisotropic character of two-point correlation function characteristic of a nematic phase.
\end{abstract}

\pacs{64.60.A-,68.35.Rh,75.70.Kw}



\maketitle

\section{Introduction}
\label{Introduction}
There is a large number of materials which present a tendency to form charge and/or magnetic stripe patterns.  These patterns can be self-organized in many different phases, such us stripes, lamellae, bubbles and others. Inhomogeneous phases of this type have been observed in systems as diverse as strongly correlated electronic systems~\cite{FradKiv2010} or ferromagnetic thin films~\cite{Pescia2010}. A common property of all these systems is that there is a certain degree of frustration, coming form the lattice structure or from 
competing interactions.  

Complex phases are generally characterized by broken symmetries manifested by the long-wavelength behavior of correlations. For instance, stripe order breaks rotational as well as translational symmetry in one direction, while crystal or bubble phases break translational invariance in all directions.  Moreover, it is possible to have a phase where the the translational symmetry is restored by the proliferation of topological defects, however the system still have orientational order. The nematic or the hexatic phases are such homogeneous phases that break rotational symmetry
while preserving translation invariance\cite{chaikin-1995}. 

In the last years there has been a renewed interest in the nematic phase, due to the fact that states with this symmetry have been observed in several highly correlated electron systems\cite{FradKiv2010} such as quantum Hall systems,  ruthenate compounds,  cuprates and Fe-based high temperature superconductors. Whether this phase is relevant to describe the interesting and exotic transport properties of such materials is still an open question. However,  there is a growing amount of data suggesting that the physics of the nematic phase could be intimately related with the non-Fermi liquid behavior of  anisotropic metal states.  

Clear evidences of an electronic nematic phase  appear  in ultra-high mobility two dimensional electron systems (2DES)  in GaAs/GaAlAs heterostructures\cite{Lilly1999,Du1999} at extremely low temperatures and moderate magnetic fields. While for huge magnetic fields the fractional quantum Hall (QH) phase dominates the physics of the first Landau level,   the nematic phase appears when the Fermi level lies near the middle of the third and higher Landau levels (smaller magnetic fields). The most evident signature of the QH nematic is the strong temperature dependence of  anisotropic transport properties\cite{Manousakis1982}.

The electronic nematic phase has  also been observed in the bilayer ruthenate  compound
Sr$_{3}$Ru$_{2}$O$_{7}$ at finite magnetic field~\cite{Borzi2007,Doh2007}. While the data 
suggests that a meta-magnetic quantum critical point can be reached by changing
the direction of the applied magnetic  field~\cite{Perry2001,Grigera2001},  transport properties of 
this system are strongly anisotropic and  its temperature dependence is very similar to
the 2DES described before.   

Other examples of stripe forming materials are the cuprates high  
$T_c$ superconductors\cite{KivFrad2003}. In addition to superconductivity, typical ingredients 
found in systems with competing interactions, such as  inhomogeneity, anisotropy, disorder and glassiness, coexist. 
The intermediate state  between the Mott insulator and the superconducting  phase is usually understood  as a spin glass  with local stripe order, called ``cluster  glass''. Although the electronic cluster glass  state exhibits no known long range order, some electronic order is always detected 
by local probes \cite{cho1992,  pana2002,  curro2006,Kohsaka2007, Lawler2010}. 
 Also, fluctuating stripes have been measured \cite{Parker2010} at the onset of the pseudo-gap state of  $\rm Bi_2Sr_2CaCu_2O_{8+x}$,  using 
spectroscopic mapping with a scanning tunneling microscope.  
In the same direction, recent measurements\cite{Daou2010} of the Nernst effect in $\rm YBa_2Cu_3O_y$, showed that the pseudo gap temperature coincides 
with the appearance of a strong in-plane anisotropy of electronic origin, compatible with the electronic nematic phase\cite{KiFrEm1998}.    

In a completely different context, stripe domains with ferromagnetic order are  observed in ultra-thin magnetic films with perpendicular
anisotropy\cite{Pescia2010,WoWuCh2005,PoVaPe2003,VaStMaPiPoPe2000,AbKaPoSa1995}. For instance, 
in monolayers of Fe grown on Cu substrates, the local magnetization shows very complex temperature 
dependent striped patterns with a bunch of topological defects.  A nematic order in these systems has been proposed by analyzing frustrated ferromagnetic models analytically \cite{BaSt2011, BaRiSt2013} as well as with  Monte-Carlo simulations\cite{CaMiStTa2006,NiSt2007}.

From a theoretical point of view,  the stripe phase was extensively studied in several systems and its origin is well understood. Different mean-field approaches correctly capture the physics of 
the stripe phase. For instance, QH stripes are correctly described by a Hartree-Fock approximation of a  
2DES  in a magnetic field with Coulomb 
interactions\cite{Koulakov1996,KouFog1996,Moessner1996}.
Hartree-Fock solutions of the Hubbard model\cite{Machida1989,Schulz1989,Zaanen1989},   slave-boson mean-field theories of
the t - J model\cite{Seibold1998, Han2001,Lorenzana2002} and even  
studies of Coulomb frustrated phase separation\cite{EmKi1993}  provide a reasonable description of the stripe phase of High $T_c$ superconductors.
In ferromagnetic thin-films, the stripe phase has been analized by means of a mean-field treatment of the 
frustrated Ising-dipolar model\cite{PiCa2007}, by elsticity of domain walls analysis of an Heisenberg hamiltonian~\cite{AbKaPoSa1995} and 
through Monte Carlo simulations~\cite{CaStTa2004}.  The hidden reason of why mean-field treatments work pretty well for the stripe phase resides in the fact that the Hamiltonian can be naturally written in terms of the stripe order parameter ({\em i.\ e.\ } local charged density or local magnetization, depending on the case).
 
All these systems, despite of the different microscopic origin, have a natural scale that dominate the stripe modulation, originated form competing interactions.
Therefore, in the spirit of the Landau theory of phase transition, it is possible to describe the stripe phase by means of a coarse grained Hamiltonian describing a system constrained to a thin shell in momentum space arond a characteristic wave vector of modulus $k_0$. 
Examples of scalar and vector order parameters behaving this way were studied in a seminal work by Brazovskii\cite{Brazovskii1985}. He showed that in systems with a spectrum of fluctuations dominated by a shell of non-zero wave vector, there is a first order phase transition at a finite temperature from an isotropic to a stripe phase, induced by field fluctuations. The main fluctuations were taken into account by implementing a self-consistent Hartree approximation, which is known to be equivalent to  the leading order term in a $1/N$ expansion of a system with $O(N)$ symmetry.

On the other hand, the nematic phase is more elusive. The nematic order parameter (as will be described in the following sections) is quadratic in the original degrees of freedom, therefore, it is essentially governed by the physics of fluctuations and cannot be captured by naive mean-field theory.
Theoretical predictions of the nematic phase are based on the study of specific interactions written in terms of the nematic order parameter. For instance, in Fermi liquids, the isotropic/nematic phase transition was studied using different techniques such as  RPA\cite{OgKiFr2001}, multidimensional bosonization\cite{BaOx2003,Lawler2006}, and Landau Fermi liquid theory\cite{CastroNeto2005, BaRe2013}, on specific models with explicit attractive quadrupole-quadrupole interaction. Moreover, in ref. \onlinecite{BaSt2007}, we have shown that in corse-grained models of the Brazovskii class, the quadrupolar interaction is naturally generated, since it is relevant  in the renormalization group sense. In particular, we have shown that in isotropic 2d models, although the stripe phase cannot exist since long wave length fluctuations diverge linearly, the nematic phase can indeed exist and it is in the Kosterlitz-Thouless universality class\cite{BaSt2009}. 

One way of describing the isotropic/nematic phase transition is to observe spontaneous Fermi surface deformations in the case of Fermi liquids, or deformations of the high temperature form factor in the case of classical systems. Provided we begin with a Hamiltonian with local polynomial interactions,  the Hartree (or Hartree-Fock for fermions) approximation provides a constant (momentum independent) renormalization to the effective Hamiltonian and then it cannot capture any deformation in momentum space. In order to have a chance to observe relevant fluctuations associated  with the isotropic-nematic phase transition it is necessary to compute corrections producing a momentum dependent self-energy . In this paper we show that the minimal approach to describe the isotropic-nematic transition in systems with isotropic competing interactions is  the self consistent screening approximation  (SCSA)\cite{Bray(a)1974,Bray(b)1974},
which is equivalent to introduce the next to leading order term in a  $1/N$ expansion of the effective Hamiltonian in a self-consistent way.   

In this work we consider the simplest Brazovskii model in two-dimensions and compute the correlation function in the SCSA. We  show that there is a critical temperature at which the system spontaneously breaks rotational symmetry, signalling the presence of an isotropic-nematic phase transition. This definitively confirms our previous claims\cite{BaSt2007,BaSt2009} based on symmetry and RG arguments within the context of a completely controlled calculation.

This paper is organized as follows:
in Section \ref{model} we introduce the model and the essential background on the nematic order parameter.
In section \ref{SCSA} we introduce the SCSA and compute, both numerically and analytically, the two-point correlation
function leading to the isotropic-nematic transition. Section \ref{conclu} is devoted to the conclusions and a 
short discussion of our results.

\section{Model Hamiltonian, stripe formation and the nematic phase}
\label{model}

We are interested in the low temperature physics of $d=2$ models with isotropic competing interactions and Ising symmetry.
Universal characteristics can be well described by a coarse-grained effective Hamiltonian written in terms of a real scalar field $\phi(\vec x)$,
of which the quadratic part in reciprocal space reads: 
\begin{equation}
H_0= \int_{\Lambda} \frac{d^{2}k}{(2\pi)^2}\;\phi(\vec k)\left(r_0 + A (k-k_0)^2+ \ldots\right) \phi(-\vec k),
\label{H0}
\end{equation}
where $r_0(T)\sim (T-T^*)$, $k=|\vec k|$ and $k_0=|\vec k_0|$ is a characteristic scale given by the competing nature of the 
microscopic interactions~\cite{SeAn1995}. $\int_{\Lambda} d^{2}k \equiv  \int_0^{2\pi} d\theta \int_{k_0-\Lambda}^{k_0+\Lambda} dk\;k$
and $\Lambda \sim \sqrt{r_0/A}$ is a cut-off where the expansion of the free energy up to quadratic order in the wave vector 
makes  sense \cite{Brazovskii1985}. 
The ``mass'' $1/A$ measures the curvature of the dispersion relation around the minimum $k_0$ and  the ellipses 
in eq.(\ref{H0}) mean higher order terms in $(k-k_0)$. 

The structure factor or, equivalently, the two-point correlation function:
\beq
 G_0(k)  = \frac{1}{ r_0+ A(k-k_0)^2} 
\label{eq.freeG}
\eeq
has a maximum at $k=k_0$ with a correlation length $\xi\sim \sqrt{A/ r_0}$. Therefore, near 
criticality ($r_0\to 0$) the physics is dominated by an annulus in reciprocal space with wave vector $k \sim k_0$ and width  $2\Lambda$.   This situation is quite similar to fermionic systems at low temperature, where $k_0$ plays the role of the Fermi momentum, and the reduction of phase space to a spherical shell (in $d=3$) centered at the Fermi momentum is  ruled by the Pauli exclusion principle. 
The question is how interactions modify this picture. 
  The simplest interaction term is given by a local quartic term of the form:
\beq \label{model2}
{\cal H}_i=v \int \left(\prod_{i=1}^4 \frac{d^2k_i}{(2\pi)^2}\right) \phi(\veck_1)\ldots\phi(\veck_4)
\delta^2(\veck_1+\veck_2+\veck_3+\veck_4), 
\eeq
where $v$ measures the interaction intensity. 
The free correlation is renormalized by the interaction term Eq. (\ref{model2}). The simplest correction
is given by the self-consistent field approximation in which the quartic term is approximated in the form
$\phi^4(\vec x)\simeq \langle \phi^2(\vec x)\rangle \phi^2(\vec x)$. In this way
the original  theory is approximated by an effective one which is quadratic in the field $\phi$ and thus can be solved  exactly. In this approximation, the correlation function $G(\veck)$ has the same free structure  of Eq. (\ref{eq.freeG})  with a renormalized $r_0\to r$, given by the self-consistent equation\cite{chaikin-1995}:
\beq \label{r.renorm}
r(T) = r_0(T) +vT \int \ddk G(\veck).
\eeq
Brazovskii showed\cite{Brazovskii1985} that the solution of this equation drives the mean field critical point to a fluctuation induced first order phase transition between an isotropic and a stripe phase characterized by a modulated order parameter $\langle \phi(\vec x) \rangle\sim \cos(k_0 x)$.  
However, long wavelength fluctuations of this modulated pattern may diverge, depending on dimensionality. In $d=3$ the 
divergence is logarithmic in the linear size of the system and there is quasi-long-ranged stripe order. However, in $d=2$ the divergence is linear implying that the system cannot order at any finite temperature. 

Then, a relevant question is about the possible existence of a homogeneous phase at intermeditate and low temperatures, like a nematic phase, which restores translation invariance but breaks rotational symmetry. 
Orientational order of this kind can be quantified by a nematic tensor order parameter given 
in  terms  of the density gradients~\cite{StBa2010}:
\begin{equation}
Q_{ij} \equiv \int d^2x \ \phi(\vec x)\left(\partial_i\partial_j-\frac{1}{2}\partial^2\delta_{ij}\right) \phi(\vec x)
\end{equation}
where $i, j = x, y$ and $\partial^2=\partial_{xx}+\partial_{yy}$ is the Laplacian in two dimensions. This tensor is symmetric and traceless, and in two dimensions it has only two independent elements which essentially represent the mean orientation of domain walls and the strength of the orientational order. In order to get some feeling of the physical content of this order parameter it is useful to write it in reciprocal space. Introducing the Fourier transforms and choosing the x-axis as the principal axis, the only relevant element of the tensor is~\cite{StBa2010}:
\begin{equation}
Q_{xx}= \int dk\, k\, \cos(2\theta)\,G(\vec k)
\label{eq.Qxx}
\end{equation}
where $k_x = k\, \cos{\theta}$, $k_y = k\, \sin{\theta}$ and $G(\vec k)$ is the structure factor of the system. Written in this way the orientational order parameter quantifies the degree of anisotropy of the domain pattern. In a completely isotropic phase, e.g. a liquid phase or a mosaic of domains with no preferential direction, the corresponding isotropy in the structure factor will be reflected in a zero value of the orientational order parameter. Therefore, any approximation such as the self consistent Hatree approximation described above, leading to a constant renormalization of the correlation  function, cannot be able to capture the physics of the nematic phase.
Note that Eq. (\ref{eq.Qxx})  acts as a filter that selects the $\cos(2\theta)$ component of the correlation function. Then, in order to have $Q_{xx}\neq 0$, the correlation function should be anisotropic. An anisotropic form leading to a nematic phase was found in Refs. \cite{BaSt2007,BaSt2009}, which has the form: 
\beq
G(\vec k)  = \frac{1}{ r+ A(k-k_0)^2+ \alpha \, k^2 \cos(2\theta)}. 
\label{eq.GNematic}
\eeq
Here, $\alpha$ is a constant which plays the role of a scalar nematic order parameter\cite{BaSt2007,BaSt2009}. If $\alpha=0$, the correlation function is isotropic and $Q_{xx}=0$. On the other hand, for $\alpha\neq 0$, the anisotropy of the correlation function gives a finite contribution to  $Q_{xx}$. 
Then, the nematic phase should be  described by an effective  quadratic  Hamiltonian of the form:
\begin{equation}
H_N= \int_{\Lambda} \frac{d^{2}k}{(2\pi)^2}\;\phi(\vec k)G(\vec k)^{-1} \phi(-\vec k)
\label{HNematic}
\end{equation}
with $G(\vec k)$ given by eq. (\ref{eq.GNematic}).
In other words,  interactions should renormalize the Hamiltonian in a momentum-dependent way. 
 This is only possible at least at two-loop approximation in a perturbative expansion. We will show in the next section that the SCSA provides the appropriate  temperature dependent  renormalization parameter $\alpha(T)$, being able to capture the isotropic-nematic phase transition. 

\section{The nematic solution in the  SCSA}
\label{SCSA}
The set of self-consistent equations for the two-point correlation function in the SCSA are (see Supplemental Material):
\begin{eqnarray}
G(\vec{k})&=&\frac{1}{G_0^{-1}(\vec{k})+\Sigma(\vec{k})} \label{G.scsa}\\ 
\Sigma(\vec{k})&=&\int\frac{d^2q}{(2\pi)^2}D(\vec{k}-\vec{q})G(\vec{q}) \label{Sigma.scsa}\\ 
D(\vec{k})&=&\frac{v}{1+v\Pi(\vec{k})} \label{D.scsa}\\ 
\Pi(\vec{k})&=&\int\frac{d^2q}{(2\pi)^2}G(\vec{k}-\vec{q})G(\vec{q}). \label{Pi.scsa}
\end{eqnarray}
A general solution of this set of equations for a particular system is a formidable task. As
discussed in section \ref{model}, we expect a solution of the form (\ref{eq.GNematic}). Thus,
our aim is to show that this set of equations admits a solution of that form, leading to an
isotropic-nematic phase transiton, and that the solution is stable below some critical temperature.

We work perturbatively at order $v^2$ in (\ref{D.scsa}), such that $D(\vec{k})=v-v^2\Pi(\vec{k})$. Then, equations
(\ref{Sigma.scsa}), (\ref{D.scsa}) and (\ref{Pi.scsa}) can be written in real space in the simple form:
\begin{eqnarray}
\label{e2}
\Sigma(\vec{x})&=&D(\vec{x})G(\vec{x})\\ 
D(\vec{x})&=&v\delta(\vec{x})-v^2\Pi(\vec{x})\\
\Pi(\vec{x})&=&G(\vec{x})^2.
\end{eqnarray}

In this way the self-energy $\Sigma(\vec{x})$ is given directly as a function of $G(\vec{x})$ :
\begin{equation}
\label{e3}
\Sigma(\vec{x})=vG(0)\delta(\vec{x})-v^2G(\vec{x})^3.
\end{equation}
Then, the system of equations that actually have to be solved is reduced to:  
\begin{eqnarray}
\label{e4}
\nonumber
G(\vec{k})&=&\frac{1}{r_0+A(k-k_0)^2+\Sigma(\vec{k})}\\ 
\Sigma(\vec{x})&=&vG(0)\delta(\vec{x})-v^2G(\vec{x})^3,
\end{eqnarray}
in which the first equation in written in reciprocal space and the second in real space because it is helpful in the 
numerical analysis. 

As discussed in section \ref{model}, the nematic solution
should be characterized by broken orientational symmetry, which in the present context means that the correlation function $G(\vec{k})$ 
will depend not only on the modulus of the wave vector $k$, but also on its orientation. Then, in an expansion of the self-energy around the 
circle of radius $k_0$, the lowest order form which is necessary to reveal an orientational symmetry breaking has the form $c_1+c_2\cos(2\theta)$,
 where $c_1$ and $c_2$ are $k$-independent coefficients and $\theta$ is the angle relative to the direction along which the rotational symmetry 
is broken. 
This implies that only the "mass" term in the correlation function will be renormalized with a $\theta$ 
dependent function. Higher order terms in the self-energy will be responsible for corrections in the values of $A$ and $k_0$ but 
do not change significantly the physics near the nematic transition.    
 
From the previous arguments we propose the following ansatz for the renormalized correlation function: 
\begin{equation}
\label{e5}
G(\vec{k})=\frac{1}{r+A(k-k_0)^2+\alpha\cos(2\theta)} .
\end{equation}
where $\vec k=(k\cos{\theta},k\sin{\theta})$ and the (approximately constant) factor $k^2\sim k_0^2$ appearing in eq. (\ref{eq.GNematic})
was absorved in $\alpha$. Note that 
the angular dependence of $G(\vec{k})$ with $\theta$ implies that when the symmetry is broken ($\alpha\neq0$) two isolated maxima appear 
over the $k_y$ axis at $k_y=k_0$ and $k_y=-k_0$, i.e., in this state the director vector of the nematic phase is along the $y$ direction in real space. 

By rescaling the parameters in the form
$A \rightarrow 1$,
$k_0 \rightarrow 1$,
$\frac{r}{Ak_0^2} \rightarrow  r$
$\frac{\alpha}{Ak_0^2} \rightarrow  \alpha$,
$\frac{v}{A^2k_0^2} \rightarrow  v$,
 the problem is expressed in a dimensionless form.

To proceed, it is useful to make a Fourier expansion of $G(k,\theta)$ given by equation (\ref{e5}). Up to second order in $\theta$ this yields :
\begin{eqnarray}
G(\vec{k})&=&\frac{1}{\sqrt{(r+(k-1)^2)^2-\alpha^2}}\left(1 -\right. \\ 
&&\left.\frac{2\alpha \cos(2\theta)}{r+(k-1)^2+\sqrt{(r+(k-1)^2)^2-\alpha^2}}\right), \nonumber
\end{eqnarray}
where the $k$-dependent Fourier coefficients are exact. Then, the correlation function in the real space can be written as:
\begin{equation}
 G(\vec{x})=H_1(x)-\alpha H_2(x)\cos(2\phi),
\end{equation}
where the angle $\phi$ is such that $\phi=\pi/2$ is along the director vector of the nematic phase and 
\begin{eqnarray}
H_1(x)&=&\int \frac{d^2k}{(2\pi)^2}\frac{e^{i\vec{k}\cdot\vec{x}}}{\sqrt{(r+(k-1)^2)^2-\alpha^2}},
\label{h1}\\ \label{h2}
H_2(x)&=&\int \frac{d^2k}{(2\pi)^2}\frac{\cos{(2\theta)}e^{i\vec{k}\cdot\vec{x}}}{\sqrt{(r+(k-1)^2)^2-\alpha^2}} \times \\ 
&&\frac{2}{r+(k-1)^2+\sqrt{(r+(k-1)^2)^2-\alpha^2}}, \nonumber
\end{eqnarray}
where $\vec{k}\cdot\vec{x}=k x \cos{\theta}$.
Then,
\begin{eqnarray}
\nonumber
G(\vec{x})^3&=&(H_1^3(x)+\frac{3}{2}\alpha^2 H_1(x) H_2^2(x))\\ 
&-&(3\alpha H_1^2(x) H_2(x)+\frac{3}{4}\alpha^3 H_2^3(x))\cos{(2\phi)}.
\end{eqnarray}

Now we are able to compute $\Sigma(\vec{k})$ by taking the inverse Fourier transform in equation (\ref{e4}). As already mentioned, we are only interested in 
$\Sigma(\vec{k}_0)$ which renormalizes the mass as a function of the angle $\theta$. 
Thus, we fixed $\vec k=\vec k_0$ and obtained $\Sigma(\vec{u})$, with $\vec{u}=\vec{k}_0/k_0$ in the rescaled variables. 

The resulting renormalization equations are: 
\begin{eqnarray}
\label{e6}
r&=&r_0+vH_1(0)-v^2\int d^2x\ (H_1^3(x)+ \nonumber \\
&&\frac{3}{2}\alpha^2H_1(x)H_2^2(x))\ e^{-ix\cos{\phi}}\\
\alpha&=&3v^2 \int d^2x\ (\alpha H_1^2(x)H_2(x)+ \nonumber \\
   &&\frac{1}{4}\alpha^3H_2^3(x))\ \cos(2\phi)\ e^{-ix\cos{\phi}}.
\end{eqnarray}
From this set of equations we can see that the solution with $\alpha=0$ is always a solution. 
A non-zero solution can be searched from the last equation which, after factoring out the $\alpha=0$ solution,  may be written in the form:
\begin{equation} \label{e7}
\alpha=\sqrt{\frac{1-3v^2\int d^2x\,H_1^2(x) H_2(x)\cos(2\phi)e^{-ix\cos{\phi}}}{\frac{3}{4}v^2\int d^2x\,H_2^3(x)\cos(2\phi)e^{-ix\cos{\phi}}}} .
\end{equation}
It is worth to note that this equation is not yet the solution of $\alpha$ since the functions $H_1(x)$ and $H_2(x)$ depend on $\alpha$ too through eqs. (\ref{h1}) and (\ref{h2}) respectively.
Nevertheless, this form is appropriate to proceed numerically.    

\subsection{Numerical results}

\begin{figure}
\begin{center}
\includegraphics[scale=0.55,angle=0]{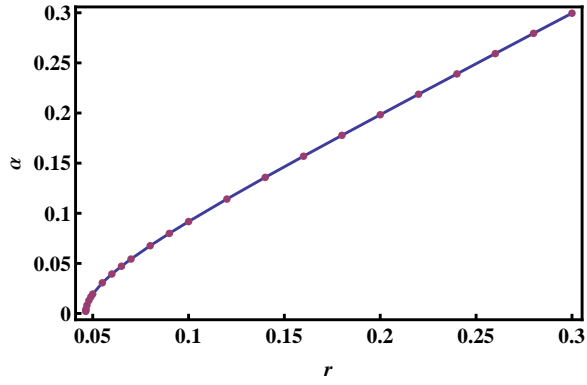}
\caption{(Color online) Numerical solution $\alpha(r)$ for $v=0.1$. 
The upper (dashed) line corresponds to the 
the solution with $\alpha \neq 0$, while the lower (full) line corresponds to the $\alpha=0$ solution.\label{1}}
\end{center}
\end{figure}

In order to solve numerically eqs. (\ref{e6}) and (\ref{e7}) it is convenient to divide the
computation in two steps: first we solve the equation for $\alpha$ for a generic set of values of $r$. In this way we build the function $\alpha(r)$ which 
is shown in Fig.\ref{1} for $v=0.1$. This
function is independent of all parameters in the system except for $v$. 

\begin{figure}
\begin{center}
\includegraphics[scale=0.55 ,angle=0]{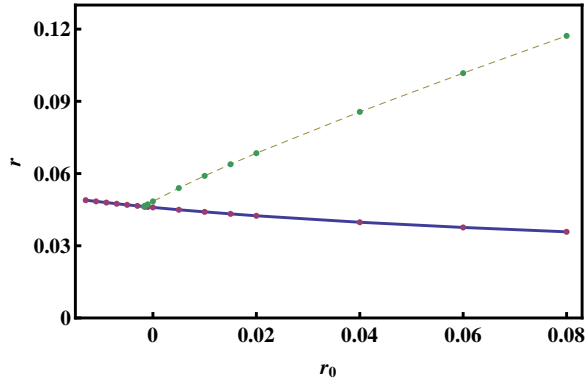}
\caption{(Color online) Numerical solution of $r$ as function of $r_0$ for $v=0.1$. \label{2}}
\end{center}
\end{figure}

Then we proceed with the computation of $r$ varying $r_0$ using equation (\ref{e6}). 
 We have to consider both cases $\alpha\neq0$ and $\alpha=0$. 
In the case of $\alpha\neq0$ we use the previously calculated function $\alpha(r)$ to solve equation (\ref{e6}), while for $\alpha=0$ 
we solve the equation for $r$ directly. 
In Fig.\ref{2} we show the solution of equation (\ref{e6}) in both cases for $v=0.1$. We see how at some critical value of $r_0=r_{0c}$ a 
bifurcation takes place and the nematic solution appears.  

\begin{figure}[ht!]
\begin{center}
\includegraphics[scale=0.55,angle=0]{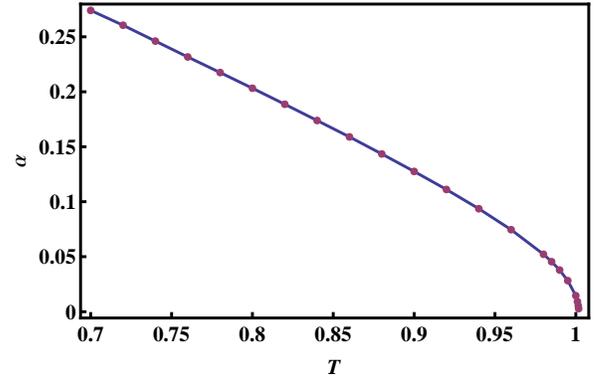}
\caption{(Color online) The nematic order parameter $\alpha$ as a function of the dimensionless temperature $T$,
 for $v=0.1$. \label{3}}
\end{center}
\end{figure}

As usual, we consider that $r_0$ depends on temperature like $r_0=1-T$, where $T$ is the dimensionless temperature. 
 In Fig.\ref{3} we show the nematic
order parameter as a function of temperature. 

\subsection{Nature of the isotropic-nematic transition in the SCSA}
\label{stab}
We have shown that the SCSA equations for the Brazovskii model admit a nematic solution ($\alpha\neq0$) as well as an isotropic one 
($\alpha=0$), as seen in Fig. \ref{2}. To establish one or the other as the thermodynamic solution we have to compare their free energies.
The free energy within the SCSA is given by (see Supplemental Material):
\begin{eqnarray}
F(T) &=& \frac{1}{2}\int\frac{d^2k}{(2\pi)^2}\ln\left(G^{-1}(\vec{k})\right)
+\frac{1}{2}\int\frac{d^2k}{(2\pi)^2}\ln\left(D^{-1}(\vec{k})\right) \nonumber \\
&-&\frac{1}{2}\int\frac{d^2k}{(2\pi)^2}\, \Sigma(\vec{k})G(\vec{k}).
\end{eqnarray}

Expanding the logarithm in the second integral to linear order in the polarization function leads to
\begin{eqnarray}
\nonumber
F(T)&=&\frac{1}{2}\int\frac{d^2k}{(2\pi)^2}\ln\left(G^{-1}(\vec{k})\right)
+\frac{v}{2}\int\frac{d^2k}{(2\pi)^2}\, \Pi(\vec{k}) \\ 
&-&\frac{1}{2}\int\frac{d^2k}{(2\pi)^2}\Sigma(\vec{k})G(\vec{k})-\frac{1}{2}\int\frac{d^2k}{(2\pi)^2}\ln{v},
\label{e9}
\end{eqnarray}
where the last term in equation (\ref{e9}) is a constant.  
The second term of the last equation is simply given by $\int\frac{d^2k}{(2\pi)^2}\Pi(\vec{k})=G^2(0)$.
The term including the self-energy can be calculated considering that the function $G(\vec{k})$ is peaked at $k_0$.
Then it is enough to consider $\Sigma(\vec{k})$ as varying only over the circle of radius $k_0$. 
The above considerations lead to a difference between the free energy of the isotropic and the nematic solutions given by:
\begin{eqnarray}
\Delta F&=&F(\alpha=0)-F(\alpha\neq0)\label{e10} \\  \nonumber
&=&\frac{1}{2}\int\frac{d^2k}{(2\pi)^2}\,\ln\left(\frac{G^{-1}_{\alpha=0}(\vec{k})}{G^{-1}_{\alpha\neq0}(\vec{k})}\right)
+\frac{v}{2}\left(G^2_{\alpha=0}(0)-G^2_{\alpha\neq0}(0)\right) \\ \nonumber
&-&\frac{1}{2}\int\frac{d^2k}{(2\pi)^2}\left(\Sigma_{\alpha=0}(\vec{u})G_{\alpha=0}(\vec{k})
- \Sigma_{\alpha\neq0}(\vec{u}) G_{\alpha\neq0}(\vec{k})\right)
\end{eqnarray}
where $\vec{u}=\vec{k}/k$.

\begin{figure}[ht!]
\begin{center}
\includegraphics[scale=0.55 ,angle=0]{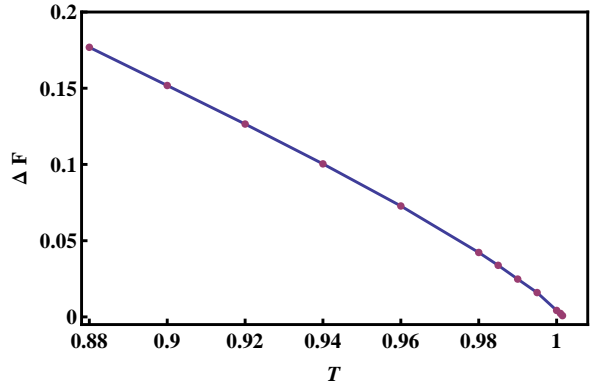}
\caption{(Color online) Free energy difference between the isotropic and the nematic solutions as 
function of dimensionless temperature for $v=0.1$. \label{4}}
\end{center}
\end{figure}

All magnitudes involved in $\Delta F$ have already been found and then a numeric evaluation is
straightforward. The results are shown in Fig.\ref{4} for $v=0.1$. We see that for $T < T_c$, corresponding to $r_0=r_{0c}$, the
difference in the free energy $\Delta F>0$, meaning that the nematic solution presents smaller free energy 
than the isotropic one. We have also confirmed this result by performing an analytical expansion of the free energy near the
critical point. This fact establishes the existence of a continuous isotropic-nematic transition in the context of the SCSA.

\subsection{Critical behavior}
\label{crit}
To analyze the behavior near the critical point in more detail it is convenient to include 
the parameters $\alpha$ and $r$ in the notations 
of the functions $H_1(x)\equiv H_1(x,r,\alpha)$ and $H_2(x) \equiv H_2(x,r,\alpha)$.
Equation (\ref{e7}) for $\alpha=0$ provides a condition to determine the critical value of 
$r=r_c$, 
and equation (\ref{e6}) allows to obtain the corresponding critical $r_0=r_{0c}$:
\begin{equation}
\nonumber
 3v^2\int d^2x\,H_1^2(x,r_c,0) H_2(x,r_c,0)\cos(2\phi)e^{-ix\cos{\phi}}=1
\end{equation}
and
\begin{equation}
 r_{0c}=r_c-vH_1(0,r_c,0)+v^2\int d^2x\,H_1^3(x,r_c,0)e^{-ix\cos{\phi}}.\\ \nonumber
\end{equation}

Expanding $H_1$ and $H_2$ to leading order in $r$ and $\alpha$ around the critical point
results in:
\begin{eqnarray}
\nonumber 
H_1(x,r,\alpha)&=&H_1(x,r_c,0)-I_{11}(x)\Delta r+I_{12}(x)\alpha^2, \\ \nonumber
H_2(x,r,\alpha)&=&H_2(x,r_c,0)-I_{21}(x)\Delta r+I_{22}(x)\alpha^2,
\end{eqnarray}
where $\Delta r=(r-r_c)$ and
\begin{eqnarray}
\nonumber
I_{11}(x)&=&\int \frac{d^2k}{(2\pi)^2}\frac{e^{i\vec{k}\cdot\vec{x}}}{(r_c+(k-1)^2)^2}\\ \nonumber
I_{12}(x)&=&\frac{1}{2}\int \frac{d^2k}{(2\pi)^2}\frac{e^{i\vec{k}\cdot\vec{x}}}{(r_c+(k-1)^2)^3}\\ \nonumber
I_{21}(x)&=&2\int \frac{d^2k}{(2\pi)^2}\frac{e^{i\vec{k}\cdot\vec{x}}\cos(2\theta)}{(r_c+(k-1)^2)^3}\\ \nonumber
I_{22}(x)&=&\frac{3}{4}\int \frac{d^2k}{(2\pi)^2}\frac{e^{i\vec{k}\cdot\vec{x}}\cos(2\theta)}{(r_c+(k-1)^2)^4}.\\ \nonumber
\end{eqnarray}
Using this expansions in equations (\ref{e6}) and (\ref{e7}) for $r$ and $\alpha$ we arrive at $\alpha=\sqrt{K\Delta r}$ with $\Delta r= m(r_0-r_{0c})$, where 
\begin{widetext}
\begin{eqnarray}
\nonumber
 K&=&\frac{\int d^2x\,e^{-ix\cos{\phi}}\cos(2\phi)\left(2H_1(x,r_c,0)H_2(x,r_c,0)I_{11}(x)+H_1(x,r_c,0)^2I_{21}(x)\right))}
 {\int d^2x\,e^{-ix\cos{\phi}}\cos(2\theta)\left(\frac{1}{4}H_2(x,r_c,0)^3+2H_1(x,r_c,0)H_2(x,r_c,0)I_{12}(x)+H_1(x,r_c,0)^2I_{22}(x)\right))},\\
 \nonumber
m^{-1}&=&1+vI_{11}(0)-vKI_{12}(0)\\ \nonumber
 &-&3v^2\int d^2x\,e^{-ix\cos{\phi}}\left(H_1(x,r_c,0)^2I_{11}(x)+KH_1(x,r_c,0)^2I_{12}(x)+\frac{K}{2}H_1(x,r_c,0)H_{2}(x,r_c,0)^2\right).
\end{eqnarray}
\end{widetext}
$K$ and $m$ are positive numbers leading to the numerical solutions shown in Fig.\ref{1} and Fig.\ref{2}. Both figures show that $\alpha(r)$ and $r(r_0)$ are increasing functions of $r$ and $r_0$ respectively. A direct conclusion
of this calculation is that $\alpha=\sqrt{mK(T_c-T)}$ near the critical temperature as is observed in Fig.\ref{3}. Then, as already found from
the analysis of the free energy, the SCSA predicts a continuous second order isotropic-nematic transition in agreement with our previous results in \onlinecite{BaSt2007,BaSt2009}. In reference \onlinecite{BaSt2009} we showed that considering angular fluctuations of the nematic order parameter
drives this transition to be of KT type.

\section{Conclusions}
\label{conclu}
The main result of this work has been to prove the existence of a isotropic-nematic continuous transition 
in a stripe forming system within a two-loop expansion in the self-consistent screening approximation. In previous
works the transition in the class of models studied here had been found by including explicitely an interaction term in
the effective Hamiltonian, ruled by symmetry considerations on the nematic phase. In that case, an additional
phenomenological interaction parameter was introduced in the spirit of a Landau expansion. Although the results were satisfactory
from a physics point of view, showing the presence of a phase transition to a nematic phase with broken rotational
symmetry in qualitative agreement with observations in several systems described by the effective Hamiltonian,
an important question remained to be answered: was it possible to obtain that interaction term from a more microscopic,
or fundamental Hamiltonian ? Here we have answered that question. The relevant interaction term can be obtained in a
perturbative expansion of the high temperature effective $\phi^4$ Hamiltonian with the usual constant interaction
parameter $v$, by going at least to two-loop order, i.e. the renormalized interaction which is responsible for the appearance
of a nematic phase is $O(v^2)$. It is now easy to see why this is so. The most simple mean field
approximations, or the one-loop Hartree approximation for the density correlations, are not able to account for a nematic phase. The physical
reason lies in the anisotropy of correlations inherent to the nematic phase. Then, any renormalization independent of
wave vector will preserve the rotation invariance of the high temperature correlations. In order that the self-energy
correction have a $\vec k$ dependence which may lead to a broken rotational symmetry, one has to go beyond Hartree approximation.  

We have found a breaking of rotational symmetry within the self-consistent screening approximation. Already at this level of
perturbation of the original Hamiltonian the equations which determine the two-point correlation function are very difficult
to solve in a closed form. Making a physically motivated assumption on the form of the simplest possible solution, 
we have solved the equations analytically near the transition and we have implemented an efficient way to solve 
them numerically away from the critical region. Our present results are in complete agreement with our previous results based
on a Landau expansion. In particular the isotropic-nematic transition is continuous with mean field critical exponent of the
nematic order parameter $\alpha \sim (T_c-T)^{1/2}$. Note that, although the SCSA already accounts for
fluctuation terms in the density field $\phi(\vec x)$, the fact that the nematic order parameter is
proportional to correlation functions implies that the present approach gives only a mean-field description of
the nematic phase. As found by us in previous work, we expect this transition to be of the 
Kosterlitz-Thouless type, i.e. a transition driven by the proliferation of topological defects, upon incorporation of relevant
fluctuation terms in the mean-field-like nematic solution. Although anisotropic phases as described in
this work have been reported in many experimental studies, as discussed in the introduction, as far as we know the experimental characterization of the isotropic-nematic phase transitions is still a big challenge.

\acknowledgments
The Brazilian agencies, {\em Funda\c c\~ao de Amparo \`a Pesquisa do Estado do Rio
de Janeiro}  (FAPERJ) and {\em Conselho Nacional de Desenvolvimento Cient\'\i
fico e Tecnol\'ogico} (CNPq) are acknowledged  for partial financial support.


\begin{thebibliography}{51}%
\makeatletter
\providecommand \@ifxundefined [1]{%
 \@ifx{#1\undefined}
}%
\providecommand \@ifnum [1]{%
 \ifnum #1\expandafter \@firstoftwo
 \else \expandafter \@secondoftwo
 \fi
}%
\providecommand \@ifx [1]{%
 \ifx #1\expandafter \@firstoftwo
 \else \expandafter \@secondoftwo
 \fi
}%
\providecommand \natexlab [1]{#1}%
\providecommand \enquote  [1]{``#1''}%
\providecommand \bibnamefont  [1]{#1}%
\providecommand \bibfnamefont [1]{#1}%
\providecommand \citenamefont [1]{#1}%
\providecommand \href@noop [0]{\@secondoftwo}%
\providecommand \href [0]{\begingroup \@sanitize@url \@href}%
\providecommand \@href[1]{\@@startlink{#1}\@@href}%
\providecommand \@@href[1]{\endgroup#1\@@endlink}%
\providecommand \@sanitize@url [0]{\catcode `\\12\catcode `\$12\catcode
  `\&12\catcode `\#12\catcode `\^12\catcode `\_12\catcode `\%12\relax}%
\providecommand \@@startlink[1]{}%
\providecommand \@@endlink[0]{}%
\providecommand \url  [0]{\begingroup\@sanitize@url \@url }%
\providecommand \@url [1]{\endgroup\@href {#1}{\urlprefix }}%
\providecommand \urlprefix  [0]{URL }%
\providecommand \Eprint [0]{\href }%
\providecommand \doibase [0]{http://dx.doi.org/}%
\providecommand \selectlanguage [0]{\@gobble}%
\providecommand \bibinfo  [0]{\@secondoftwo}%
\providecommand \bibfield  [0]{\@secondoftwo}%
\providecommand \translation [1]{[#1]}%
\providecommand \BibitemOpen [0]{}%
\providecommand \bibitemStop [0]{}%
\providecommand \bibitemNoStop [0]{.\EOS\space}%
\providecommand \EOS [0]{\spacefactor3000\relax}%
\providecommand \BibitemShut  [1]{\csname bibitem#1\endcsname}%
\let\auto@bib@innerbib\@empty
\bibitem [{\citenamefont {Fradkin}\ \emph {et~al.}(2010)\citenamefont
  {Fradkin}, \citenamefont {Kivelson}, \citenamefont {Lawler}, \citenamefont
  {Eisenstein},\ and\ \citenamefont {Mackenzie}}]{FradKiv2010}%
  \BibitemOpen
  \bibfield  {author} {\bibinfo {author} {\bibfnamefont {E.}~\bibnamefont
  {Fradkin}}, \bibinfo {author} {\bibfnamefont {S.~A.}\ \bibnamefont
  {Kivelson}}, \bibinfo {author} {\bibfnamefont {M.~J.}\ \bibnamefont
  {Lawler}}, \bibinfo {author} {\bibfnamefont {J.~P.}\ \bibnamefont
  {Eisenstein}}, \ and\ \bibinfo {author} {\bibfnamefont {A.~P.}\ \bibnamefont
  {Mackenzie}},\ }\href {\doibase 10.1146/annurev-conmatphys-070909-103925}
  {\bibfield  {journal} {\bibinfo  {journal} {Annual Review of Condensed Matter
  Physics}\ }\textbf {\bibinfo {volume} {1}},\ \bibinfo {pages} {153} (\bibinfo
  {year} {2010})},\ \Eprint
  {http://arxiv.org/abs/http://www.annualreviews.org/doi/pdf/10.1146/annurev-c%
onmatphys-070909-103925}
  {http://www.annualreviews.org/doi/pdf/10.1146/annurev-conmatphys-070909-1039%
25} \BibitemShut {NoStop}%
\bibitem [{\citenamefont {Saratz}\ \emph {et~al.}(2010)\citenamefont {Saratz},
  \citenamefont {Lichtenberger}, \citenamefont {Portmann}, \citenamefont
  {Ramsperger}, \citenamefont {Vindigni},\ and\ \citenamefont
  {Pescia}}]{Pescia2010}%
  \BibitemOpen
  \bibfield  {author} {\bibinfo {author} {\bibfnamefont {N.}~\bibnamefont
  {Saratz}}, \bibinfo {author} {\bibfnamefont {A.}~\bibnamefont
  {Lichtenberger}}, \bibinfo {author} {\bibfnamefont {O.}~\bibnamefont
  {Portmann}}, \bibinfo {author} {\bibfnamefont {U.}~\bibnamefont
  {Ramsperger}}, \bibinfo {author} {\bibfnamefont {A.}~\bibnamefont
  {Vindigni}}, \ and\ \bibinfo {author} {\bibfnamefont {D.}~\bibnamefont
  {Pescia}},\ }\href {\doibase 10.1103/PhysRevLett.104.077203} {\bibfield
  {journal} {\bibinfo  {journal} {Phys. Rev. Lett.}\ }\textbf {\bibinfo
  {volume} {104}},\ \bibinfo {pages} {077203} (\bibinfo {year}
  {2010})}\BibitemShut {NoStop}%
\bibitem [{\citenamefont {Chaikin}\ and\ \citenamefont
  {Lubensky}(1995)}]{chaikin-1995}%
  \BibitemOpen
  \bibfield  {author} {\bibinfo {author} {\bibfnamefont {P.~M.}\ \bibnamefont
  {Chaikin}}\ and\ \bibinfo {author} {\bibfnamefont {T.~C.}\ \bibnamefont
  {Lubensky}},\ }\href@noop {} {\emph {\bibinfo {title} {{Principles of
  Condensed Matter Physics}}}}\ (\bibinfo  {publisher} {Cambridge University
  Press},\ \bibinfo {address} {Cambridge, UK},\ \bibinfo {year}
  {1995})\BibitemShut {NoStop}%
\bibitem [{\citenamefont {Lilly}\ \emph {et~al.}(1999)\citenamefont {Lilly},
  \citenamefont {Cooper}, \citenamefont {Eisenstein}, \citenamefont
  {Pfeiffer},\ and\ \citenamefont {West}}]{Lilly1999}%
  \BibitemOpen
  \bibfield  {author} {\bibinfo {author} {\bibfnamefont {M.~P.}\ \bibnamefont
  {Lilly}}, \bibinfo {author} {\bibfnamefont {K.~B.}\ \bibnamefont {Cooper}},
  \bibinfo {author} {\bibfnamefont {J.~P.}\ \bibnamefont {Eisenstein}},
  \bibinfo {author} {\bibfnamefont {L.~N.}\ \bibnamefont {Pfeiffer}}, \ and\
  \bibinfo {author} {\bibfnamefont {K.~W.}\ \bibnamefont {West}},\ }\href
  {\doibase 10.1103/PhysRevLett.82.394} {\bibfield  {journal} {\bibinfo
  {journal} {Phys. Rev. Lett.}\ }\textbf {\bibinfo {volume} {82}},\ \bibinfo
  {pages} {394} (\bibinfo {year} {1999})}\BibitemShut {NoStop}%
\bibitem [{\citenamefont {Du}\ \emph {et~al.}(1999)\citenamefont {Du},
  \citenamefont {Tsui}, \citenamefont {Stormer}, \citenamefont {Pfeiffer},
  \citenamefont {Baldwin},\ and\ \citenamefont {West}}]{Du1999}%
  \BibitemOpen
  \bibfield  {author} {\bibinfo {author} {\bibfnamefont {R.}~\bibnamefont
  {Du}}, \bibinfo {author} {\bibfnamefont {D.}~\bibnamefont {Tsui}}, \bibinfo
  {author} {\bibfnamefont {H.}~\bibnamefont {Stormer}}, \bibinfo {author}
  {\bibfnamefont {L.}~\bibnamefont {Pfeiffer}}, \bibinfo {author}
  {\bibfnamefont {K.}~\bibnamefont {Baldwin}}, \ and\ \bibinfo {author}
  {\bibfnamefont {K.}~\bibnamefont {West}},\ }\href {\doibase
  http://dx.doi.org/10.1016/S0038-1098(98)00578-X} {\bibfield  {journal}
  {\bibinfo  {journal} {Solid State Communications}\ }\textbf {\bibinfo
  {volume} {109}},\ \bibinfo {pages} {389 } (\bibinfo {year}
  {1999})}\BibitemShut {NoStop}%
\bibitem [{\citenamefont {Fradkin}\ \emph {et~al.}(2000)\citenamefont
  {Fradkin}, \citenamefont {Kivelson}, \citenamefont {Manousakis},\ and\
  \citenamefont {Nho}}]{Manousakis1982}%
  \BibitemOpen
  \bibfield  {author} {\bibinfo {author} {\bibfnamefont {E.}~\bibnamefont
  {Fradkin}}, \bibinfo {author} {\bibfnamefont {S.~A.}\ \bibnamefont
  {Kivelson}}, \bibinfo {author} {\bibfnamefont {E.}~\bibnamefont
  {Manousakis}}, \ and\ \bibinfo {author} {\bibfnamefont {K.}~\bibnamefont
  {Nho}},\ }\href {\doibase 10.1103/PhysRevLett.84.1982} {\bibfield  {journal}
  {\bibinfo  {journal} {Phys. Rev. Lett.}\ }\textbf {\bibinfo {volume} {84}},\
  \bibinfo {pages} {1982} (\bibinfo {year} {2000})}\BibitemShut {NoStop}%
\bibitem [{\citenamefont {Borzi}\ \emph {et~al.}(2007)\citenamefont {Borzi},
  \citenamefont {Grigera}, \citenamefont {Farrell}, \citenamefont {Perry},
  \citenamefont {Lister}, \citenamefont {Lee}, \citenamefont {Tennant},
  \citenamefont {Maeno},\ and\ \citenamefont {Mackenzie}}]{Borzi2007}%
  \BibitemOpen
  \bibfield  {author} {\bibinfo {author} {\bibfnamefont {R.~A.}\ \bibnamefont
  {Borzi}}, \bibinfo {author} {\bibfnamefont {S.~A.}\ \bibnamefont {Grigera}},
  \bibinfo {author} {\bibfnamefont {J.}~\bibnamefont {Farrell}}, \bibinfo
  {author} {\bibfnamefont {R.~S.}\ \bibnamefont {Perry}}, \bibinfo {author}
  {\bibfnamefont {S.~J.~S.}\ \bibnamefont {Lister}}, \bibinfo {author}
  {\bibfnamefont {S.~L.}\ \bibnamefont {Lee}}, \bibinfo {author} {\bibfnamefont
  {D.~A.}\ \bibnamefont {Tennant}}, \bibinfo {author} {\bibfnamefont
  {Y.}~\bibnamefont {Maeno}}, \ and\ \bibinfo {author} {\bibfnamefont {A.~P.}\
  \bibnamefont {Mackenzie}},\ }\href {\doibase 10.1126/science.1134796}
  {\bibfield  {journal} {\bibinfo  {journal} {Science}\ }\textbf {\bibinfo
  {volume} {315}},\ \bibinfo {pages} {214} (\bibinfo {year} {2007})},\ \Eprint
  {http://arxiv.org/abs/http://www.sciencemag.org/content/315/5809/214.full.pd%
f} {http://www.sciencemag.org/content/315/5809/214.full.pdf} \BibitemShut
  {NoStop}%
\bibitem [{\citenamefont {Doh}\ \emph {et~al.}(2007)\citenamefont {Doh},
  \citenamefont {Kim},\ and\ \citenamefont {Ahn}}]{Doh2007}%
  \BibitemOpen
  \bibfield  {author} {\bibinfo {author} {\bibfnamefont {H.}~\bibnamefont
  {Doh}}, \bibinfo {author} {\bibfnamefont {Y.~B.}\ \bibnamefont {Kim}}, \ and\
  \bibinfo {author} {\bibfnamefont {K.~H.}\ \bibnamefont {Ahn}},\ }\href
  {\doibase 10.1103/PhysRevLett.98.126407} {\bibfield  {journal} {\bibinfo
  {journal} {Phys. Rev. Lett.}\ }\textbf {\bibinfo {volume} {98}},\ \bibinfo
  {pages} {126407} (\bibinfo {year} {2007})}\BibitemShut {NoStop}%
\bibitem [{\citenamefont {Perry}\ \emph {et~al.}(2001)\citenamefont {Perry},
  \citenamefont {Galvin}, \citenamefont {Grigera}, \citenamefont {Capogna},
  \citenamefont {Schofield}, \citenamefont {Mackenzie}, \citenamefont {Chiao},
  \citenamefont {Julian}, \citenamefont {Ikeda}, \citenamefont {Nakatsuji},
  \citenamefont {Maeno},\ and\ \citenamefont {Pfleiderer}}]{Perry2001}%
  \BibitemOpen
  \bibfield  {author} {\bibinfo {author} {\bibfnamefont {R.~S.}\ \bibnamefont
  {Perry}}, \bibinfo {author} {\bibfnamefont {L.~M.}\ \bibnamefont {Galvin}},
  \bibinfo {author} {\bibfnamefont {S.~A.}\ \bibnamefont {Grigera}}, \bibinfo
  {author} {\bibfnamefont {L.}~\bibnamefont {Capogna}}, \bibinfo {author}
  {\bibfnamefont {A.~J.}\ \bibnamefont {Schofield}}, \bibinfo {author}
  {\bibfnamefont {A.~P.}\ \bibnamefont {Mackenzie}}, \bibinfo {author}
  {\bibfnamefont {M.}~\bibnamefont {Chiao}}, \bibinfo {author} {\bibfnamefont
  {S.~R.}\ \bibnamefont {Julian}}, \bibinfo {author} {\bibfnamefont {S.~I.}\
  \bibnamefont {Ikeda}}, \bibinfo {author} {\bibfnamefont {S.}~\bibnamefont
  {Nakatsuji}}, \bibinfo {author} {\bibfnamefont {Y.}~\bibnamefont {Maeno}}, \
  and\ \bibinfo {author} {\bibfnamefont {C.}~\bibnamefont {Pfleiderer}},\
  }\href {\doibase 10.1103/PhysRevLett.86.2661} {\bibfield  {journal} {\bibinfo
   {journal} {Phys. Rev. Lett.}\ }\textbf {\bibinfo {volume} {86}},\ \bibinfo
  {pages} {2661} (\bibinfo {year} {2001})}\BibitemShut {NoStop}%
\bibitem [{\citenamefont {Grigera}\ \emph {et~al.}(2001)\citenamefont
  {Grigera}, \citenamefont {Perry}, \citenamefont {Schofield}, \citenamefont
  {Chiao}, \citenamefont {Julian}, \citenamefont {Lonzarich}, \citenamefont
  {Ikeda}, \citenamefont {Maeno}, \citenamefont {Millis},\ and\ \citenamefont
  {Mackenzie}}]{Grigera2001}%
  \BibitemOpen
  \bibfield  {author} {\bibinfo {author} {\bibfnamefont {S.~A.}\ \bibnamefont
  {Grigera}}, \bibinfo {author} {\bibfnamefont {R.~S.}\ \bibnamefont {Perry}},
  \bibinfo {author} {\bibfnamefont {A.~J.}\ \bibnamefont {Schofield}}, \bibinfo
  {author} {\bibfnamefont {M.}~\bibnamefont {Chiao}}, \bibinfo {author}
  {\bibfnamefont {S.~R.}\ \bibnamefont {Julian}}, \bibinfo {author}
  {\bibfnamefont {G.~G.}\ \bibnamefont {Lonzarich}}, \bibinfo {author}
  {\bibfnamefont {S.~I.}\ \bibnamefont {Ikeda}}, \bibinfo {author}
  {\bibfnamefont {Y.}~\bibnamefont {Maeno}}, \bibinfo {author} {\bibfnamefont
  {A.~J.}\ \bibnamefont {Millis}}, \ and\ \bibinfo {author} {\bibfnamefont
  {A.~P.}\ \bibnamefont {Mackenzie}},\ }\href {\doibase
  10.1126/science.1063539} {\bibfield  {journal} {\bibinfo  {journal}
  {Science}\ }\textbf {\bibinfo {volume} {294}},\ \bibinfo {pages} {329}
  (\bibinfo {year} {2001})},\ \Eprint
  {http://arxiv.org/abs/http://www.sciencemag.org/content/294/5541/329.full.pd%
f} {http://www.sciencemag.org/content/294/5541/329.full.pdf} \BibitemShut
  {NoStop}%
\bibitem [{\citenamefont {Kivelson}\ \emph {et~al.}(2003)\citenamefont
  {Kivelson}, \citenamefont {Bindloss}, \citenamefont {Fradkin}, \citenamefont
  {Oganesyan}, \citenamefont {Tranquada}, \citenamefont {Kapitulnik},\ and\
  \citenamefont {Howald}}]{KivFrad2003}%
  \BibitemOpen
  \bibfield  {author} {\bibinfo {author} {\bibfnamefont {S.~A.}\ \bibnamefont
  {Kivelson}}, \bibinfo {author} {\bibfnamefont {I.~P.}\ \bibnamefont
  {Bindloss}}, \bibinfo {author} {\bibfnamefont {E.}~\bibnamefont {Fradkin}},
  \bibinfo {author} {\bibfnamefont {V.}~\bibnamefont {Oganesyan}}, \bibinfo
  {author} {\bibfnamefont {J.~M.}\ \bibnamefont {Tranquada}}, \bibinfo {author}
  {\bibfnamefont {A.}~\bibnamefont {Kapitulnik}}, \ and\ \bibinfo {author}
  {\bibfnamefont {C.}~\bibnamefont {Howald}},\ }\href {\doibase
  10.1103/RevModPhys.75.1201} {\bibfield  {journal} {\bibinfo  {journal} {Rev.
  Mod. Phys.}\ }\textbf {\bibinfo {volume} {75}},\ \bibinfo {pages} {1201}
  (\bibinfo {year} {2003})}\BibitemShut {NoStop}%
\bibitem [{\citenamefont {Cho}\ \emph {et~al.}(1992)\citenamefont {Cho},
  \citenamefont {Borsa}, \citenamefont {Johnston},\ and\ \citenamefont
  {Torgeson}}]{cho1992}%
  \BibitemOpen
  \bibfield  {author} {\bibinfo {author} {\bibfnamefont {J.~H.}\ \bibnamefont
  {Cho}}, \bibinfo {author} {\bibfnamefont {F.}~\bibnamefont {Borsa}}, \bibinfo
  {author} {\bibfnamefont {D.~C.}\ \bibnamefont {Johnston}}, \ and\ \bibinfo
  {author} {\bibfnamefont {D.~R.}\ \bibnamefont {Torgeson}},\ }\href {\doibase
  10.1103/PhysRevB.46.3179} {\bibfield  {journal} {\bibinfo  {journal} {Phys.
  Rev. B}\ }\textbf {\bibinfo {volume} {46}},\ \bibinfo {pages} {3179}
  (\bibinfo {year} {1992})}\BibitemShut {NoStop}%
\bibitem [{\citenamefont {Panagopoulos}\ \emph {et~al.}(2002)\citenamefont
  {Panagopoulos}, \citenamefont {Tallon}, \citenamefont {Rainford},
  \citenamefont {Xiang}, \citenamefont {Cooper},\ and\ \citenamefont
  {Scott}}]{pana2002}%
  \BibitemOpen
  \bibfield  {author} {\bibinfo {author} {\bibfnamefont {C.}~\bibnamefont
  {Panagopoulos}}, \bibinfo {author} {\bibfnamefont {J.~L.}\ \bibnamefont
  {Tallon}}, \bibinfo {author} {\bibfnamefont {B.~D.}\ \bibnamefont
  {Rainford}}, \bibinfo {author} {\bibfnamefont {T.}~\bibnamefont {Xiang}},
  \bibinfo {author} {\bibfnamefont {J.~R.}\ \bibnamefont {Cooper}}, \ and\
  \bibinfo {author} {\bibfnamefont {C.~A.}\ \bibnamefont {Scott}},\ }\href
  {\doibase 10.1103/PhysRevB.66.064501} {\bibfield  {journal} {\bibinfo
  {journal} {Phys. Rev. B}\ }\textbf {\bibinfo {volume} {66}},\ \bibinfo
  {pages} {064501} (\bibinfo {year} {2002})}\BibitemShut {NoStop}%
\bibitem [{\citenamefont {Grafe}\ \emph {et~al.}(2006)\citenamefont {Grafe},
  \citenamefont {Curro}, \citenamefont {H\"ucker},\ and\ \citenamefont
  {B\"uchner}}]{curro2006}%
  \BibitemOpen
  \bibfield  {author} {\bibinfo {author} {\bibfnamefont {H.-J.}\ \bibnamefont
  {Grafe}}, \bibinfo {author} {\bibfnamefont {N.~J.}\ \bibnamefont {Curro}},
  \bibinfo {author} {\bibfnamefont {M.}~\bibnamefont {H\"ucker}}, \ and\
  \bibinfo {author} {\bibfnamefont {B.}~\bibnamefont {B\"uchner}},\ }\href
  {\doibase 10.1103/PhysRevLett.96.017002} {\bibfield  {journal} {\bibinfo
  {journal} {Phys. Rev. Lett.}\ }\textbf {\bibinfo {volume} {96}},\ \bibinfo
  {pages} {017002} (\bibinfo {year} {2006})}\BibitemShut {NoStop}%
\bibitem [{\citenamefont {Kohsaka}\ \emph {et~al.}(2007)\citenamefont
  {Kohsaka}, \citenamefont {Taylor}, \citenamefont {Fujita}, \citenamefont
  {Schmidt}, \citenamefont {Lupien}, \citenamefont {Hanaguri}, \citenamefont
  {Azuma}, \citenamefont {Takano}, \citenamefont {Eisaki}, \citenamefont
  {Takagi}, \citenamefont {Uchida},\ and\ \citenamefont {Davis}}]{Kohsaka2007}%
  \BibitemOpen
  \bibfield  {author} {\bibinfo {author} {\bibfnamefont {Y.}~\bibnamefont
  {Kohsaka}}, \bibinfo {author} {\bibfnamefont {C.}~\bibnamefont {Taylor}},
  \bibinfo {author} {\bibfnamefont {K.}~\bibnamefont {Fujita}}, \bibinfo
  {author} {\bibfnamefont {A.}~\bibnamefont {Schmidt}}, \bibinfo {author}
  {\bibfnamefont {C.}~\bibnamefont {Lupien}}, \bibinfo {author} {\bibfnamefont
  {T.}~\bibnamefont {Hanaguri}}, \bibinfo {author} {\bibfnamefont
  {M.}~\bibnamefont {Azuma}}, \bibinfo {author} {\bibfnamefont
  {M.}~\bibnamefont {Takano}}, \bibinfo {author} {\bibfnamefont
  {H.}~\bibnamefont {Eisaki}}, \bibinfo {author} {\bibfnamefont
  {H.}~\bibnamefont {Takagi}}, \bibinfo {author} {\bibfnamefont
  {S.}~\bibnamefont {Uchida}}, \ and\ \bibinfo {author} {\bibfnamefont {J.~C.}\
  \bibnamefont {Davis}},\ }\href {\doibase 10.1126/science.1138584} {\bibfield
  {journal} {\bibinfo  {journal} {Science}\ }\textbf {\bibinfo {volume}
  {315}},\ \bibinfo {pages} {1380} (\bibinfo {year} {2007})},\ \Eprint
  {http://arxiv.org/abs/http://www.sciencemag.org/content/315/5817/1380.full.p%
df} {http://www.sciencemag.org/content/315/5817/1380.full.pdf} \BibitemShut
  {NoStop}%
\bibitem [{\citenamefont {Lawler}\ \emph {et~al.}(2010)\citenamefont {Lawler},
  \citenamefont {Fujita}, \citenamefont {Lee}, \citenamefont {Schmidt},
  \citenamefont {Kohsaka}, \citenamefont {Kim}, \citenamefont {Eisaki},
  \citenamefont {Uchida}, \citenamefont {Davis}, \citenamefont {Sethna},\ and\
  \citenamefont {Kim}}]{Lawler2010}%
  \BibitemOpen
  \bibfield  {author} {\bibinfo {author} {\bibfnamefont {M.~J.}\ \bibnamefont
  {Lawler}}, \bibinfo {author} {\bibfnamefont {K.}~\bibnamefont {Fujita}},
  \bibinfo {author} {\bibfnamefont {J.}~\bibnamefont {Lee}}, \bibinfo {author}
  {\bibfnamefont {A.~R.}\ \bibnamefont {Schmidt}}, \bibinfo {author}
  {\bibfnamefont {Y.}~\bibnamefont {Kohsaka}}, \bibinfo {author} {\bibfnamefont
  {C.~K.}\ \bibnamefont {Kim}}, \bibinfo {author} {\bibfnamefont
  {H.}~\bibnamefont {Eisaki}}, \bibinfo {author} {\bibfnamefont
  {S.}~\bibnamefont {Uchida}}, \bibinfo {author} {\bibfnamefont {J.~C.}\
  \bibnamefont {Davis}}, \bibinfo {author} {\bibfnamefont {J.~P.}\ \bibnamefont
  {Sethna}}, \ and\ \bibinfo {author} {\bibfnamefont {E.-A.}\ \bibnamefont
  {Kim}},\ }\href {http://dx.doi.org/10.1038/nature09169} {\bibfield  {journal}
  {\bibinfo  {journal} {Nature}\ }\textbf {\bibinfo {volume} {466}},\ \bibinfo
  {pages} {347} (\bibinfo {year} {2010})}\BibitemShut {NoStop}%
\bibitem [{\citenamefont {Parker}\ \emph {et~al.}(2010)\citenamefont {Parker},
  \citenamefont {Aynajian}, \citenamefont {da~Silva~Neto}, \citenamefont
  {Pushp}, \citenamefont {Ono}, \citenamefont {Wen}, \citenamefont {Xu},
  \citenamefont {Gu},\ and\ \citenamefont {Yazdani}}]{Parker2010}%
  \BibitemOpen
  \bibfield  {author} {\bibinfo {author} {\bibfnamefont {C.~V.}\ \bibnamefont
  {Parker}}, \bibinfo {author} {\bibfnamefont {P.}~\bibnamefont {Aynajian}},
  \bibinfo {author} {\bibfnamefont {E.~H.}\ \bibnamefont {da~Silva~Neto}},
  \bibinfo {author} {\bibfnamefont {A.}~\bibnamefont {Pushp}}, \bibinfo
  {author} {\bibfnamefont {S.}~\bibnamefont {Ono}}, \bibinfo {author}
  {\bibfnamefont {J.}~\bibnamefont {Wen}}, \bibinfo {author} {\bibfnamefont
  {Z.}~\bibnamefont {Xu}}, \bibinfo {author} {\bibfnamefont {G.}~\bibnamefont
  {Gu}}, \ and\ \bibinfo {author} {\bibfnamefont {A.}~\bibnamefont {Yazdani}},\
  }\href {http://dx.doi.org/10.1038/nature09597} {\bibfield  {journal}
  {\bibinfo  {journal} {Nature}\ }\textbf {\bibinfo {volume} {468}},\ \bibinfo
  {pages} {677} (\bibinfo {year} {2010})}\BibitemShut {NoStop}%
\bibitem [{\citenamefont {Daou}\ \emph {et~al.}(2010)\citenamefont {Daou},
  \citenamefont {Chang}, \citenamefont {LeBoeuf}, \citenamefont
  {Cyr-Choiniere}, \citenamefont {Laliberte}, \citenamefont {Doiron-Leyraud},
  \citenamefont {Ramshaw}, \citenamefont {Liang}, \citenamefont {Bonn},
  \citenamefont {Hardy},\ and\ \citenamefont {Taillefer}}]{Daou2010}%
  \BibitemOpen
  \bibfield  {author} {\bibinfo {author} {\bibfnamefont {R.}~\bibnamefont
  {Daou}}, \bibinfo {author} {\bibfnamefont {J.}~\bibnamefont {Chang}},
  \bibinfo {author} {\bibfnamefont {D.}~\bibnamefont {LeBoeuf}}, \bibinfo
  {author} {\bibfnamefont {O.}~\bibnamefont {Cyr-Choiniere}}, \bibinfo {author}
  {\bibfnamefont {F.}~\bibnamefont {Laliberte}}, \bibinfo {author}
  {\bibfnamefont {N.}~\bibnamefont {Doiron-Leyraud}}, \bibinfo {author}
  {\bibfnamefont {B.~J.}\ \bibnamefont {Ramshaw}}, \bibinfo {author}
  {\bibfnamefont {R.}~\bibnamefont {Liang}}, \bibinfo {author} {\bibfnamefont
  {D.~A.}\ \bibnamefont {Bonn}}, \bibinfo {author} {\bibfnamefont {W.~N.}\
  \bibnamefont {Hardy}}, \ and\ \bibinfo {author} {\bibfnamefont
  {L.}~\bibnamefont {Taillefer}},\ }\href
  {http://dx.doi.org/10.1038/nature08716} {\bibfield  {journal} {\bibinfo
  {journal} {Nature}\ }\textbf {\bibinfo {volume} {463}},\ \bibinfo {pages}
  {519} (\bibinfo {year} {2010})}\BibitemShut {NoStop}%
\bibitem [{\citenamefont {Kivelson}\ \emph {et~al.}(1998)\citenamefont
  {Kivelson}, \citenamefont {Fradkin},\ and\ \citenamefont
  {Emery}}]{KiFrEm1998}%
  \BibitemOpen
  \bibfield  {author} {\bibinfo {author} {\bibfnamefont {S.~A.}\ \bibnamefont
  {Kivelson}}, \bibinfo {author} {\bibfnamefont {E.}~\bibnamefont {Fradkin}}, \
  and\ \bibinfo {author} {\bibfnamefont {V.~J.}\ \bibnamefont {Emery}},\
  }\href@noop {} {\bibfield  {journal} {\bibinfo  {journal} {Nature}\ }\textbf
  {\bibinfo {volume} {393}},\ \bibinfo {pages} {550} (\bibinfo {year}
  {1998})}\BibitemShut {NoStop}%
\bibitem [{\citenamefont {Won}\ \emph {et~al.}(2005)\citenamefont {Won},
  \citenamefont {Wu}, \citenamefont {Choi}, \citenamefont {Kim}, \citenamefont
  {Scholl}, \citenamefont {Doran}, \citenamefont {Owens}, \citenamefont {Wu},
  \citenamefont {Jin},\ and\ \citenamefont {Qiu}}]{WoWuCh2005}%
  \BibitemOpen
  \bibfield  {author} {\bibinfo {author} {\bibfnamefont {C.}~\bibnamefont
  {Won}}, \bibinfo {author} {\bibfnamefont {Y.~Z.}\ \bibnamefont {Wu}},
  \bibinfo {author} {\bibfnamefont {J.}~\bibnamefont {Choi}}, \bibinfo {author}
  {\bibfnamefont {W.}~\bibnamefont {Kim}}, \bibinfo {author} {\bibfnamefont
  {A.}~\bibnamefont {Scholl}}, \bibinfo {author} {\bibfnamefont
  {A.}~\bibnamefont {Doran}}, \bibinfo {author} {\bibfnamefont
  {T.}~\bibnamefont {Owens}}, \bibinfo {author} {\bibfnamefont
  {J.}~\bibnamefont {Wu}}, \bibinfo {author} {\bibfnamefont {X.~F.}\
  \bibnamefont {Jin}}, \ and\ \bibinfo {author} {\bibfnamefont {Z.~Q.}\
  \bibnamefont {Qiu}},\ }\href@noop {} {\bibfield  {journal} {\bibinfo
  {journal} {Phys. Rev. B}\ }\textbf {\bibinfo {volume} {71}},\ \bibinfo
  {pages} {224429} (\bibinfo {year} {2005})}\BibitemShut {NoStop}%
\bibitem [{\citenamefont {Portmann}\ \emph {et~al.}(2003)\citenamefont
  {Portmann}, \citenamefont {Vaterlaus},\ and\ \citenamefont
  {Pescia}}]{PoVaPe2003}%
  \BibitemOpen
  \bibfield  {author} {\bibinfo {author} {\bibfnamefont {O.}~\bibnamefont
  {Portmann}}, \bibinfo {author} {\bibfnamefont {A.}~\bibnamefont {Vaterlaus}},
  \ and\ \bibinfo {author} {\bibfnamefont {D.}~\bibnamefont {Pescia}},\
  }\href@noop {} {\bibfield  {journal} {\bibinfo  {journal} {Nature}\ }\textbf
  {\bibinfo {volume} {422}},\ \bibinfo {pages} {701} (\bibinfo {year}
  {2003})}\BibitemShut {NoStop}%
\bibitem [{\citenamefont {Vaterlaus}\ \emph {et~al.}(2000)\citenamefont
  {Vaterlaus}, \citenamefont {Stamm}, \citenamefont {Maier}, \citenamefont
  {Pini}, \citenamefont {Politi},\ and\ \citenamefont
  {Pescia}}]{VaStMaPiPoPe2000}%
  \BibitemOpen
  \bibfield  {author} {\bibinfo {author} {\bibfnamefont {A.}~\bibnamefont
  {Vaterlaus}}, \bibinfo {author} {\bibfnamefont {C.}~\bibnamefont {Stamm}},
  \bibinfo {author} {\bibfnamefont {U.}~\bibnamefont {Maier}}, \bibinfo
  {author} {\bibfnamefont {M.~G.}\ \bibnamefont {Pini}}, \bibinfo {author}
  {\bibfnamefont {P.}~\bibnamefont {Politi}}, \ and\ \bibinfo {author}
  {\bibfnamefont {D.}~\bibnamefont {Pescia}},\ }\href@noop {} {\bibfield
  {journal} {\bibinfo  {journal} {Phys. Rev. Lett.}\ }\textbf {\bibinfo
  {volume} {84}},\ \bibinfo {pages} {2247} (\bibinfo {year}
  {2000})}\BibitemShut {NoStop}%
\bibitem [{\citenamefont {Abanov}\ \emph {et~al.}(1995)\citenamefont {Abanov},
  \citenamefont {Kalatsky}, \citenamefont {Pokrovsky},\ and\ \citenamefont
  {Saslow}}]{AbKaPoSa1995}%
  \BibitemOpen
  \bibfield  {author} {\bibinfo {author} {\bibfnamefont {A.}~\bibnamefont
  {Abanov}}, \bibinfo {author} {\bibfnamefont {V.}~\bibnamefont {Kalatsky}},
  \bibinfo {author} {\bibfnamefont {V.~L.}\ \bibnamefont {Pokrovsky}}, \ and\
  \bibinfo {author} {\bibfnamefont {W.~M.}\ \bibnamefont {Saslow}},\
  }\href@noop {} {\bibfield  {journal} {\bibinfo  {journal} {Phys. Rev. B}\
  }\textbf {\bibinfo {volume} {51}},\ \bibinfo {pages} {1023} (\bibinfo {year}
  {1995})}\BibitemShut {NoStop}%
\bibitem [{\citenamefont {Barci}\ and\ \citenamefont
  {Stariolo}(2011)}]{BaSt2011}%
  \BibitemOpen
  \bibfield  {author} {\bibinfo {author} {\bibfnamefont {D.~G.}\ \bibnamefont
  {Barci}}\ and\ \bibinfo {author} {\bibfnamefont {D.~A.}\ \bibnamefont
  {Stariolo}},\ }\href {\doibase 10.1103/PhysRevB.84.094439} {\bibfield
  {journal} {\bibinfo  {journal} {Phys. Rev. B}\ }\textbf {\bibinfo {volume}
  {84}},\ \bibinfo {pages} {094439} (\bibinfo {year} {2011})}\BibitemShut
  {NoStop}%
\bibitem [{\citenamefont {Barci}\ \emph {et~al.}(2013)\citenamefont {Barci},
  \citenamefont {Ribeiro},\ and\ \citenamefont {Stariolo}}]{BaRiSt2013}%
  \BibitemOpen
  \bibfield  {author} {\bibinfo {author} {\bibfnamefont {D.~G.}\ \bibnamefont
  {Barci}}, \bibinfo {author} {\bibfnamefont {L.}~\bibnamefont {Ribeiro}}, \
  and\ \bibinfo {author} {\bibfnamefont {D.~A.}\ \bibnamefont {Stariolo}},\
  }\href {\doibase 10.1103/PhysRevE.87.062119} {\bibfield  {journal} {\bibinfo
  {journal} {Phys. Rev. E}\ }\textbf {\bibinfo {volume} {87}},\ \bibinfo
  {pages} {062119} (\bibinfo {year} {2013})}\BibitemShut {NoStop}%
\bibitem [{\citenamefont {Cannas}\ \emph {et~al.}(2006)\citenamefont {Cannas},
  \citenamefont {Michelon}, \citenamefont {Stariolo},\ and\ \citenamefont
  {Tamarit}}]{CaMiStTa2006}%
  \BibitemOpen
  \bibfield  {author} {\bibinfo {author} {\bibfnamefont {S.~A.}\ \bibnamefont
  {Cannas}}, \bibinfo {author} {\bibfnamefont {M.~F.}\ \bibnamefont
  {Michelon}}, \bibinfo {author} {\bibfnamefont {D.~A.}\ \bibnamefont
  {Stariolo}}, \ and\ \bibinfo {author} {\bibfnamefont {F.~A.}\ \bibnamefont
  {Tamarit}},\ }\href {\doibase 10.1103/PhysRevB.73.184425} {\bibfield
  {journal} {\bibinfo  {journal} {Phys. Rev. B}\ }\textbf {\bibinfo {volume}
  {73}},\ \bibinfo {pages} {184425} (\bibinfo {year} {2006})}\BibitemShut
  {NoStop}%
\bibitem [{\citenamefont {Nicolao}\ and\ \citenamefont
  {Stariolo}(2007)}]{NiSt2007}%
  \BibitemOpen
  \bibfield  {author} {\bibinfo {author} {\bibfnamefont {L.}~\bibnamefont
  {Nicolao}}\ and\ \bibinfo {author} {\bibfnamefont {D.~A.}\ \bibnamefont
  {Stariolo}},\ }\href {\doibase 10.1103/PhysRevB.76.054453} {\bibfield
  {journal} {\bibinfo  {journal} {Phys. Rev. B .}\ }\textbf {\bibinfo {volume}
  {76}},\ \bibinfo {eid} {054453} (\bibinfo {year} {2007})}\BibitemShut
  {NoStop}%
\bibitem [{\citenamefont {Koulakov}\ \emph {et~al.}(1996)\citenamefont
  {Koulakov}, \citenamefont {Fogler},\ and\ \citenamefont
  {Shklovskii}}]{Koulakov1996}%
  \BibitemOpen
  \bibfield  {author} {\bibinfo {author} {\bibfnamefont {A.~A.}\ \bibnamefont
  {Koulakov}}, \bibinfo {author} {\bibfnamefont {M.~M.}\ \bibnamefont
  {Fogler}}, \ and\ \bibinfo {author} {\bibfnamefont {B.~I.}\ \bibnamefont
  {Shklovskii}},\ }\href {\doibase 10.1103/PhysRevLett.76.499} {\bibfield
  {journal} {\bibinfo  {journal} {Phys. Rev. Lett.}\ }\textbf {\bibinfo
  {volume} {76}},\ \bibinfo {pages} {499} (\bibinfo {year} {1996})}\BibitemShut
  {NoStop}%
\bibitem [{\citenamefont {Fogler}\ \emph {et~al.}(1996)\citenamefont {Fogler},
  \citenamefont {Koulakov},\ and\ \citenamefont {Shklovskii}}]{KouFog1996}%
  \BibitemOpen
  \bibfield  {author} {\bibinfo {author} {\bibfnamefont {M.~M.}\ \bibnamefont
  {Fogler}}, \bibinfo {author} {\bibfnamefont {A.~A.}\ \bibnamefont
  {Koulakov}}, \ and\ \bibinfo {author} {\bibfnamefont {B.~I.}\ \bibnamefont
  {Shklovskii}},\ }\href {\doibase 10.1103/PhysRevB.54.1853} {\bibfield
  {journal} {\bibinfo  {journal} {Phys. Rev. B}\ }\textbf {\bibinfo {volume}
  {54}},\ \bibinfo {pages} {1853} (\bibinfo {year} {1996})}\BibitemShut
  {NoStop}%
\bibitem [{\citenamefont {Moessner}\ and\ \citenamefont
  {Chalker}(1996)}]{Moessner1996}%
  \BibitemOpen
  \bibfield  {author} {\bibinfo {author} {\bibfnamefont {R.}~\bibnamefont
  {Moessner}}\ and\ \bibinfo {author} {\bibfnamefont {J.~T.}\ \bibnamefont
  {Chalker}},\ }\href {\doibase 10.1103/PhysRevB.54.5006} {\bibfield  {journal}
  {\bibinfo  {journal} {Phys. Rev. B}\ }\textbf {\bibinfo {volume} {54}},\
  \bibinfo {pages} {5006} (\bibinfo {year} {1996})}\BibitemShut {NoStop}%
\bibitem [{\citenamefont {Machida}(1989)}]{Machida1989}%
  \BibitemOpen
  \bibfield  {author} {\bibinfo {author} {\bibfnamefont {K.}~\bibnamefont
  {Machida}},\ }\href {\doibase http://dx.doi.org/10.1016/0921-4534(89)90316-X}
  {\bibfield  {journal} {\bibinfo  {journal} {Physica C: Superconductivity}\
  }\textbf {\bibinfo {volume} {158}},\ \bibinfo {pages} {192 } (\bibinfo {year}
  {1989})}\BibitemShut {NoStop}%
\bibitem [{\citenamefont {{Schulz, H.J.}}(1989)}]{Schulz1989}%
  \BibitemOpen
  \bibfield  {author} {\bibinfo {author} {\bibnamefont {{Schulz, H.J.}}},\
  }\href {\doibase 10.1051/jphys:0198900500180283300} {\bibfield  {journal}
  {\bibinfo  {journal} {J. Phys. France}\ }\textbf {\bibinfo {volume} {50}},\
  \bibinfo {pages} {2833} (\bibinfo {year} {1989})}\BibitemShut {NoStop}%
\bibitem [{\citenamefont {Zaanen}\ and\ \citenamefont
  {Gunnarsson}(1989)}]{Zaanen1989}%
  \BibitemOpen
  \bibfield  {author} {\bibinfo {author} {\bibfnamefont {J.}~\bibnamefont
  {Zaanen}}\ and\ \bibinfo {author} {\bibfnamefont {O.}~\bibnamefont
  {Gunnarsson}},\ }\href {\doibase 10.1103/PhysRevB.40.7391} {\bibfield
  {journal} {\bibinfo  {journal} {Phys. Rev. B}\ }\textbf {\bibinfo {volume}
  {40}},\ \bibinfo {pages} {7391} (\bibinfo {year} {1989})}\BibitemShut
  {NoStop}%
\bibitem [{\citenamefont {Seibold}\ \emph {et~al.}(1998)\citenamefont
  {Seibold}, \citenamefont {Sigmund},\ and\ \citenamefont
  {Hizhnyakov}}]{Seibold1998}%
  \BibitemOpen
  \bibfield  {author} {\bibinfo {author} {\bibfnamefont {G.}~\bibnamefont
  {Seibold}}, \bibinfo {author} {\bibfnamefont {E.}~\bibnamefont {Sigmund}}, \
  and\ \bibinfo {author} {\bibfnamefont {V.}~\bibnamefont {Hizhnyakov}},\
  }\href {\doibase 10.1103/PhysRevB.57.6937} {\bibfield  {journal} {\bibinfo
  {journal} {Phys. Rev. B}\ }\textbf {\bibinfo {volume} {57}},\ \bibinfo
  {pages} {6937} (\bibinfo {year} {1998})}\BibitemShut {NoStop}%
\bibitem [{\citenamefont {HAN}\ \emph {et~al.}(2001)\citenamefont {HAN},
  \citenamefont {WANG},\ and\ \citenamefont {LEE}}]{Han2001}%
  \BibitemOpen
  \bibfield  {author} {\bibinfo {author} {\bibfnamefont {J.}~\bibnamefont
  {HAN}}, \bibinfo {author} {\bibfnamefont {Q.-H.}\ \bibnamefont {WANG}}, \
  and\ \bibinfo {author} {\bibfnamefont {D.-H.}\ \bibnamefont {LEE}},\ }\href
  {\doibase 10.1142/S021797920100468X} {\bibfield  {journal} {\bibinfo
  {journal} {International Journal of Modern Physics B}\ }\textbf {\bibinfo
  {volume} {15}},\ \bibinfo {pages} {1117} (\bibinfo {year} {2001})},\ \Eprint
  {http://arxiv.org/abs/http://www.worldscientific.com/doi/pdf/10.1142/S021797%
920100468X} {http://www.worldscientific.com/doi/pdf/10.1142/S021797920100468X}
  \BibitemShut {NoStop}%
\bibitem [{\citenamefont {Lorenzana}\ and\ \citenamefont
  {Seibold}(2002)}]{Lorenzana2002}%
  \BibitemOpen
  \bibfield  {author} {\bibinfo {author} {\bibfnamefont {J.}~\bibnamefont
  {Lorenzana}}\ and\ \bibinfo {author} {\bibfnamefont {G.}~\bibnamefont
  {Seibold}},\ }\href {\doibase 10.1103/PhysRevLett.89.136401} {\bibfield
  {journal} {\bibinfo  {journal} {Phys. Rev. Lett.}\ }\textbf {\bibinfo
  {volume} {89}},\ \bibinfo {pages} {136401} (\bibinfo {year}
  {2002})}\BibitemShut {NoStop}%
\bibitem [{\citenamefont {Emery}\ and\ \citenamefont
  {Kivelson}(1993)}]{EmKi1993}%
  \BibitemOpen
  \bibfield  {author} {\bibinfo {author} {\bibfnamefont {V.}~\bibnamefont
  {Emery}}\ and\ \bibinfo {author} {\bibfnamefont {S.}~\bibnamefont
  {Kivelson}},\ }\href {\doibase 10.1016/0921-4534(93)90581-A} {\bibfield
  {journal} {\bibinfo  {journal} {Physica C: Superconductivity}\ }\textbf
  {\bibinfo {volume} {209}},\ \bibinfo {pages} {597 } (\bibinfo {year}
  {1993})}\BibitemShut {NoStop}%
\bibitem [{\citenamefont {Pigh\'{\i}n}\ and\ \citenamefont
  {Cannas}(2007)}]{PiCa2007}%
  \BibitemOpen
  \bibfield  {author} {\bibinfo {author} {\bibfnamefont {S.~A.}\ \bibnamefont
  {Pigh\'{\i}n}}\ and\ \bibinfo {author} {\bibfnamefont {S.~A.}\ \bibnamefont
  {Cannas}},\ }\href {\doibase 10.1103/PhysRevB.75.224433} {\bibfield
  {journal} {\bibinfo  {journal} {Phys. Rev. B}\ }\textbf {\bibinfo {volume}
  {75}},\ \bibinfo {eid} {224433} (\bibinfo {year} {2007})}\BibitemShut
  {NoStop}%
\bibitem [{\citenamefont {Cannas}\ \emph {et~al.}(2004)\citenamefont {Cannas},
  \citenamefont {Stariolo},\ and\ \citenamefont {Tamarit}}]{CaStTa2004}%
  \BibitemOpen
  \bibfield  {author} {\bibinfo {author} {\bibfnamefont {S.~A.}\ \bibnamefont
  {Cannas}}, \bibinfo {author} {\bibfnamefont {D.~A.}\ \bibnamefont
  {Stariolo}}, \ and\ \bibinfo {author} {\bibfnamefont {F.~A.}\ \bibnamefont
  {Tamarit}},\ }\href {\doibase 10.1103/PhysRevB.69.092409} {\bibfield
  {journal} {\bibinfo  {journal} {Phys. Rev. B}\ }\textbf {\bibinfo {volume}
  {69}},\ \bibinfo {pages} {092409} (\bibinfo {year} {2004})}\BibitemShut
  {NoStop}%
\bibitem [{\citenamefont {Brazovskii}(1985)}]{Brazovskii1985}%
  \BibitemOpen
  \bibfield  {author} {\bibinfo {author} {\bibfnamefont {S.~A.}\ \bibnamefont
  {Brazovskii}},\ }\href@noop {} {\bibfield  {journal} {\bibinfo  {journal}
  {Sov.\ Phys.\ JETP}\ }\textbf {\bibinfo {volume} {41}},\ \bibinfo {pages}
  {85} (\bibinfo {year} {1985})}\BibitemShut {NoStop}%
\bibitem [{\citenamefont {Oganesyan}\ \emph {et~al.}(2001)\citenamefont
  {Oganesyan}, \citenamefont {Kivelson},\ and\ \citenamefont
  {Fradkin}}]{OgKiFr2001}%
  \BibitemOpen
  \bibfield  {author} {\bibinfo {author} {\bibfnamefont {V.}~\bibnamefont
  {Oganesyan}}, \bibinfo {author} {\bibfnamefont {S.~A.}\ \bibnamefont
  {Kivelson}}, \ and\ \bibinfo {author} {\bibfnamefont {E.}~\bibnamefont
  {Fradkin}},\ }\href {\doibase 10.1103/PhysRevB.64.195109} {\bibfield
  {journal} {\bibinfo  {journal} {Phys. Rev. B}\ }\textbf {\bibinfo {volume}
  {64}},\ \bibinfo {pages} {195109} (\bibinfo {year} {2001})}\BibitemShut
  {NoStop}%
\bibitem [{\citenamefont {Barci}\ and\ \citenamefont {Oxman}(2003)}]{BaOx2003}%
  \BibitemOpen
  \bibfield  {author} {\bibinfo {author} {\bibfnamefont {D.~G.}\ \bibnamefont
  {Barci}}\ and\ \bibinfo {author} {\bibfnamefont {L.~E.}\ \bibnamefont
  {Oxman}},\ }\href {\doibase 10.1103/PhysRevB.67.205108} {\bibfield  {journal}
  {\bibinfo  {journal} {Phys. Rev. B}\ }\textbf {\bibinfo {volume} {67}},\
  \bibinfo {pages} {205108} (\bibinfo {year} {2003})}\BibitemShut {NoStop}%
\bibitem [{\citenamefont {Lawler}\ \emph {et~al.}(2006)\citenamefont {Lawler},
  \citenamefont {Barci}, \citenamefont {Fern\'andez}, \citenamefont {Fradkin},\
  and\ \citenamefont {Oxman}}]{Lawler2006}%
  \BibitemOpen
  \bibfield  {author} {\bibinfo {author} {\bibfnamefont {M.~J.}\ \bibnamefont
  {Lawler}}, \bibinfo {author} {\bibfnamefont {D.~G.}\ \bibnamefont {Barci}},
  \bibinfo {author} {\bibfnamefont {V.}~\bibnamefont {Fern\'andez}}, \bibinfo
  {author} {\bibfnamefont {E.}~\bibnamefont {Fradkin}}, \ and\ \bibinfo
  {author} {\bibfnamefont {L.}~\bibnamefont {Oxman}},\ }\href {\doibase
  10.1103/PhysRevB.73.085101} {\bibfield  {journal} {\bibinfo  {journal} {Phys.
  Rev. B}\ }\textbf {\bibinfo {volume} {73}},\ \bibinfo {pages} {085101}
  (\bibinfo {year} {2006})}\BibitemShut {NoStop}%
\bibitem [{\citenamefont {Nilsson}\ and\ \citenamefont
  {Castro~Neto}(2005)}]{CastroNeto2005}%
  \BibitemOpen
  \bibfield  {author} {\bibinfo {author} {\bibfnamefont {J.}~\bibnamefont
  {Nilsson}}\ and\ \bibinfo {author} {\bibfnamefont {A.~H.}\ \bibnamefont
  {Castro~Neto}},\ }\href {\doibase 10.1103/PhysRevB.72.195104} {\bibfield
  {journal} {\bibinfo  {journal} {Phys. Rev. B}\ }\textbf {\bibinfo {volume}
  {72}},\ \bibinfo {pages} {195104} (\bibinfo {year} {2005})}\BibitemShut
  {NoStop}%
\bibitem [{\citenamefont {Barci}\ and\ \citenamefont {Reyes}(2013)}]{BaRe2013}%
  \BibitemOpen
  \bibfield  {author} {\bibinfo {author} {\bibfnamefont {D.~G.}\ \bibnamefont
  {Barci}}\ and\ \bibinfo {author} {\bibfnamefont {D.}~\bibnamefont {Reyes}},\
  }\href {\doibase 10.1103/PhysRevB.87.075147} {\bibfield  {journal} {\bibinfo
  {journal} {Phys. Rev. B}\ }\textbf {\bibinfo {volume} {87}},\ \bibinfo
  {pages} {075147} (\bibinfo {year} {2013})}\BibitemShut {NoStop}%
\bibitem [{\citenamefont {Barci}\ and\ \citenamefont
  {Stariolo}(2007)}]{BaSt2007}%
  \BibitemOpen
  \bibfield  {author} {\bibinfo {author} {\bibfnamefont {D.~G.}\ \bibnamefont
  {Barci}}\ and\ \bibinfo {author} {\bibfnamefont {D.~A.}\ \bibnamefont
  {Stariolo}},\ }\href {\doibase 10.1103/PhysRevLett.98.200604} {\bibfield
  {journal} {\bibinfo  {journal} {Physical Review Letters}\ }\textbf {\bibinfo
  {volume} {98}},\ \bibinfo {eid} {200604} (\bibinfo {year}
  {2007})}\BibitemShut {NoStop}%
\bibitem [{\citenamefont {Barci}\ and\ \citenamefont
  {Stariolo}(2009)}]{BaSt2009}%
  \BibitemOpen
  \bibfield  {author} {\bibinfo {author} {\bibfnamefont {D.~G.}\ \bibnamefont
  {Barci}}\ and\ \bibinfo {author} {\bibfnamefont {D.~A.}\ \bibnamefont
  {Stariolo}},\ }\href {\doibase 10.1103/PhysRevB.79.075437} {\bibfield
  {journal} {\bibinfo  {journal} {Physical Review B}\ }\textbf {\bibinfo
  {volume} {79}},\ \bibinfo {eid} {075437} (\bibinfo {year}
  {2009})}\BibitemShut {NoStop}%
\bibitem [{\citenamefont {Bray}(1974{\natexlab{a}})}]{Bray(a)1974}%
  \BibitemOpen
  \bibfield  {author} {\bibinfo {author} {\bibfnamefont {A.~J.}\ \bibnamefont
  {Bray}},\ }\href {\doibase 10.1103/PhysRevLett.32.1413} {\bibfield  {journal}
  {\bibinfo  {journal} {Phys. Rev. Lett.}\ }\textbf {\bibinfo {volume} {32}},\
  \bibinfo {pages} {1413} (\bibinfo {year} {1974}{\natexlab{a}})}\BibitemShut
  {NoStop}%
\bibitem [{\citenamefont {Bray}(1974{\natexlab{b}})}]{Bray(b)1974}%
  \BibitemOpen
  \bibfield  {author} {\bibinfo {author} {\bibfnamefont {A.}~\bibnamefont
  {Bray}},\ }\href {\doibase 10.1007/BF01019476} {\bibfield  {journal}
  {\bibinfo  {journal} {Journal of Statistical Physics}\ }\textbf {\bibinfo
  {volume} {11}},\ \bibinfo {pages} {29} (\bibinfo {year}
  {1974}{\natexlab{b}})}\BibitemShut {NoStop}%
\bibitem [{\citenamefont {Seul}\ and\ \citenamefont
  {Andelman}(1995)}]{SeAn1995}%
  \BibitemOpen
  \bibfield  {author} {\bibinfo {author} {\bibfnamefont {M.}~\bibnamefont
  {Seul}}\ and\ \bibinfo {author} {\bibfnamefont {D.}~\bibnamefont
  {Andelman}},\ }\href@noop {} {\bibfield  {journal} {\bibinfo  {journal}
  {Science}\ }\textbf {\bibinfo {volume} {267}},\ \bibinfo {pages} {476}
  (\bibinfo {year} {1995})}\BibitemShut {NoStop}%
\bibitem [{\citenamefont {Stariolo}\ and\ \citenamefont
  {Barci}(2010)}]{StBa2010}%
  \BibitemOpen
  \bibfield  {author} {\bibinfo {author} {\bibfnamefont {D.~A.}\ \bibnamefont
  {Stariolo}}\ and\ \bibinfo {author} {\bibfnamefont {D.~G.}\ \bibnamefont
  {Barci}},\ }\href@noop {} {\bibfield  {journal} {\bibinfo  {journal} {J.
  Phys.: Conf. Series}\ }\textbf {\bibinfo {volume} {246}},\ \bibinfo {pages}
  {012021} (\bibinfo {year} {2010})}\BibitemShut {NoStop}%
\end{thebibliography}

%

\end{document}